\documentclass[11pt]{article}\textwidth 6.5in\textheight 9in
\usepackage{amssymb}\usepackage[colorlinks]{hyperref}\usepackage{color}
	\usepackage{stmaryrd}\usepackage{mathrsfs}
	\usepackage{graphicx}
	\usepackage{amsmath}
\usepackage{caption}

\begin{document}
\setlength{\captionmargin}{27pt}
\newcommand\hreff[1]{\href {http://#1} {\small http://#1}}
\newcommand\trm[1]{{\bf\em #1}} \newcommand\emm[1]{{\ensuremath{#1}}}
\newcommand\prf{\paragraph{Proof.}}\newcommand\qed{\hfill\emm\blacksquare}

\newtheorem{thr}{Theorem} 
\newtheorem{lmm}{Lemma}
\newtheorem{cor}{Corollary}
\newtheorem{con}{Conjecture} 
\newtheorem{prp}{Proposition}

\newtheorem{blk}{Block}
\newtheorem{dff}{Definition}
\newtheorem{asm}{Assumption}
\newtheorem{rmk}{Remark}
\newtheorem{clm}{Claim}
\newtheorem{example}{Example}

\newcommand{\ab}{a\!b}
\newcommand{\yx}{y\!x}
\newcommand{\yux}{y\!\underline{x}}

\newcommand\floor[1]{{\lfloor#1\rfloor}}\newcommand\ceil[1]{{\lceil#1\rceil}}

\newcommand{\lea}{<^+}
\newcommand{\gea}{>^+}
\newcommand{\eqa}{=^+}

\newcommand{\lel}{<^{\log}}
\newcommand{\gel}{>^{\log}}
\newcommand{\eql}{=^{\log}}

\newcommand{\lem}{\stackrel{\ast}{<}}
\newcommand{\gem}{\stackrel{\ast}{>}}
\newcommand{\eqm}{\stackrel{\ast}{=}}

\newcommand\edf{{\,\stackrel{\mbox{\tiny def}}=\,}}
\newcommand\edl{{\,\stackrel{\mbox{\tiny def}}\leq\,}}
\newcommand\then{\Rightarrow}

\newcommand\C{\mathbf{C}} 

\renewcommand\chi{\mathcal{H}}
\newcommand\km{{\mathbf {km}}}\renewcommand\t{{\mathbf {t}}}
\newcommand\KM{{\mathbf {KM}}}\newcommand\m{{\mathbf {m}}}
\newcommand\md{{\mathbf {m}_{\mathbf{d}}}}\newcommand\mT{{\mathbf {m}_{\mathbf{T}}}}
\newcommand\K{{\mathbf K}} \newcommand\I{{\mathbf I}}

\newcommand\II{\hat{\mathbf I}}
\newcommand\Kd{{\mathbf{Kd}}} \newcommand\KT{{\mathbf{KT}}} 
\renewcommand\d{{\mathbf d}} 
\newcommand\D{{\mathbf D}}

\newcommand\w{{\mathbf w}}\newcommand\Ks{\mathbf{Ks}} 
\newcommand\Cs{\mathbf{Cs}} \newcommand\q{{\mathbf q}}
\newcommand\E{{\mathbf E}} \newcommand\St{{\mathbf S}}
\newcommand\M{{\mathbf M}}\newcommand\Q{{\mathbf Q}}
\newcommand\ch{{\mathcal H}} \renewcommand\l{\tau}
\newcommand\tb{{\mathbf t}} \renewcommand\L{{\mathbf L}}
\newcommand\bb{{\mathbf {bb}}}\newcommand\Km{{\mathbf {Km}}}
\renewcommand\q{{\mathbf q}}\newcommand\J{{\mathbf J}}
\newcommand\z{\mathbf{z}}
\renewcommand\i{\mathbf{i}}

\newcommand\B{\mathbf{bb}}\newcommand\f{\mathbf{f}}
\newcommand\hd{\mathbf{0'}} \newcommand\T{{\mathbf T}}
\newcommand\R{\mathbb{R}}\renewcommand\Q{\mathbb{Q}}
\newcommand\N{\mathbb{N}}\newcommand\BT{\Sigma}
\newcommand\FS{\BT^*}\newcommand\IS{\BT^\infty}
\newcommand\FIS{\BT^{*\infty}}
\renewcommand\S{\mathcal{C}}\newcommand\ST{\mathcal{S}}
\newcommand\UM{\nu_0}\newcommand\EN{\mathcal{W}}

\newcommand{\supp}{\mathrm{Supp}}

\newcommand\lenum{\lbrack\!\lbrack}
\newcommand\renum{\rbrack\!\rbrack}

\newcommand\h{\mathbf{h}}
\renewcommand\qed{\hfill\emm\square}

\title{\vspace*{-3pc} Outliers, Dynamics, and the Independence Postulate}
\author {Samuel Epstein\footnote{JP Theory Group. samepst@jptheorygroup.org}}

\maketitle
\begin{abstract}
We show that outliers occur almost surely in computable dynamics over infinite sequences.  Ever greater outliers can be found as the number of visited states increases. We show the Independence Postulate explains how outliers are found in the physical world. We generalize the outliers theorem to uncomputable sampling methods.
\end{abstract}
\section{Introduction}

An outlier is an observation whose value lies outside the set of values considered likely according to some hypothesis (usually one based on other observations); an isolated point. In the realm of algorithmic information theory, outliers are modeled using the deficiency of randomness. The deficiency of randomness of a number $a\in\N$ with respect to a computable probability measure $p$ over $\N$ is $\d(a|p)=-\log p(a)-\K(a|\langle p\rangle)$. the randomness deficiency of an infinite sequence $\alpha\in\IS$ with respect to a computable probability measure $P$ over $\IS$ is $\D(\alpha|P) = \sup_n -\log P(\alpha[0..n])-\K(\alpha[0..n]|\langle P\rangle)$. The term $\K$ is the prefix free Kolmogorov complexity.

The definition of a randomness deficiency is a very useful mathematical definition, and can be used to show the ubiquity of outliers in different constructs. A sampling method is a probabilistic program that when given $n$, outputs with probability 1, $n$ unique elements, either numbers or infinite sequences. In \cite{Epstein21}, it was proven that sampling methods produce outliers. In  \cite{Epstein22}, it was proven that for ergodic dynamical systems, ever greater outlying states occur with diminishing measure. In this paper, we show that arbitrary (potentially non-ergodic) computable dynamical systems over $\IS$ produce outliers, where as more states are visited, more and greater outliers are guaranteed to occur. 

A computable dynamical system $(\lambda,\delta)$ consists of a computable starting state probability $\lambda$ over $\IS$ and a computable transition function $\delta:\IS\rightarrow\IS$. We assume that $\lambda$-a.e. starting states are aperiodic.\\

\noindent\textbf{Theorem.} \textit{
There exists $d\in \N$, where for computable probability $\mu$ over $\IS$ and computable dynamics $(\lambda,\delta)$, for $\lambda$-a.e. starting states $\alpha\in\IS$, there exists $s_\alpha\in\N$, where among the first $2^m$ states visited, for any $n<m$, there are at least $2^n$ states $\beta$ with $\D(\beta|\mu) > m-n-d\log m-s_\alpha$. Furthermore, for the smallest such $s_\alpha$, $\E_{\alpha\sim\lambda}\left[s_\alpha-O(\log s_\alpha)\right]< \K(\lambda,\delta)+O(\log \K(\mu))$.}\\

The proof technique is two stages. First is to prove that certain finite sets of natural numbers or infinite sequences have high mutual information, $\I$, with the halting sequence. This is represented in Theorems \ref{thr:stochsub} and \ref{thr:subsetinf}. The second step is to use conservation properties of $\I$, shown in Theorem \ref{thr:coninfo}, to achieve the main theorem of the paper. In fact, as shown in Section \ref{sec:outdyn}, a stronger version of the above theorem is proven, generalizing to arbitrary (i.e. uncomputable) dynamics. The above theorem holds if the (potentially infinite) encoding of the dynamical system has finite mutual information with the halting sequence. This generalization is made possible due the two step process described above. Another generalization of the above theorem, detailed in the Discussion, is for the probability measure $\mu$ to be arbitrary, i.e. uncomputable.

The above approach is compatible with the Independence Postulate. The Independence Postulate (\textbf{IP}), \cite{Levin84,Levin13}, is an unprovable inequality on the information content shared between two sequences. \textbf{IP} is a finitary Church Turing Thesis, postulating that certain infinite and finite sequences cannot be found in nature, i.e. have high “physical addresses”. In this paper we show that \textbf{IP} explains why outliers are found in the physical world. The approach in \textbf{IP} is different from that of the main theorem of this paper and \cite{Epstein21,Epstein22}. While the latter shows that computable constructs produce outliers with high probability, the former states that individual sequences without outliers have high addresses, i.e. are hard to find nature. Both methods start with proving certain sequences have mutual information with the halting sequence (though in some theorems in \cite{Epstein22}, better bounds are achieved with more direct proofs). In Section \ref{sec:ip} the following statement is derived from \textbf{IP}.\\

\noindent\textbf{Statement.} \textit{For computable probability $p$ over $\N$ and $\tau \in \N^\N$ and $\tau(n)$ equal to the first $2^n$ unique numbers of $\tau$, let $s_{\tau,p}=\sup_n(n-\max_{a \in \tau(n)} \d(a|p) - 3\K(n) )$. Then for every physical address $k$ of $\tau$, $s_{\tau,p} \lel k + O(\log\K(p))$.
}\\

The sequence $\tau$ can be seen to be a sequence of observations, each encoded by a number, $\tau[i] \in \N$. Furthermore, $\tau$ is assumed to have an infinite number of unique numbers. Section \ref{sec:ip} also details known relationships between \textbf{IP} and the halting problem, Peano Arithmetic, and inductive inference \cite{Levin13}. We recommend reading Section \ref{sec:ip} first. Section \ref{sec:unsamp} improves upon the outliers theorem in \cite{Epstein21}, showing that (potentially uncomputable) sampling methods that have finite mutual information with the halting sequence will produce outliers.
\section{Conventions}
\label{sec:conv}
We use $\N$, $\Q$, $\R$, $\BT$, $\FS$, and $\IS$ to represent natural numbers, rational numbers, reals, bits, finite strings, and infinite strings. The removal of the last bit of a string is denoted by $(p0^-){=}(p1^-){=}p$, for $p\in\FS$. We use $\FIS$ to denote $\FS{\cup}\IS$, the set of finite and infinite strings.  For $x\in \FIS$, $y\in \FIS$, we say $x\sqsubseteq y$ if $x=y$ or $x\in\FS$ and $y=xz$ for some $z\in \FIS$. Also $x\sqsubset y$ if $x\sqsubseteq y$ and $x\neq y$. The indicator function of a mathematical statement $A$ is denoted by $[A]$, where if $A$ is true then $[A]=1$, otherwise $[A]=0$. For sets $Z$ of infinite strings, $Z_{n}=\{\alpha[0..n]\,{:}\,\alpha\,{\in}\,Z\}$ and $\langle Z\rangle=\langle Z_{ 1}\rangle\langle Z_{ 2}\rangle\langle Z_{ 3}\rangle\dots$. 

	As is typical of the field of algorithmic information theory, the theorems in this paper are relative to a fixed universal  machine, and therefore their statements are only relative up to additive and logarithmic precision. For positive real functions $f$ the terms  ${\lea}f$, ${\gea}f$, ${\eqa}f$ represent ${<}f{+}O(1)$, ${>}f{-}O(1)$, and ${=}f{\pm}O(1)$, respectively. In addition ${\lem}f$, ${\gem}f$ denote $<f/O(1)$, $>f/O(1)$. The terms ${\eqm}f$  denotes ${\lem}f$ and ${\gem}f$. For nonnegative real function $f$, the terms ${\lel}f$, ${\gel} f$, ${\eql}f$ represent the terms ${<}f{+}O(\log(f{+}1))$, ${>}f{-}O(\log(f{+}1))$, and ${=}f{\pm}O(\log(f{+}1))$, respectively. A discrete measure is a nonnegative function $Q:\N\rightarrow \R_{\geq 0}$ over natural numbers. The support of a measure $Q$ is the set of all elements whose $Q$ value is positive, with $\supp(Q) = \{a\,{:}\,Q(a)>0\}$. A measure is elementary if its support is finite and its range is a subset of $\Q$. We say that $Q$ is a probability measure if $\sum_aQ(a)\,{=}\,1$.

$T_y(x)$ is the output of algorithm $T$ (or $\perp$ if it does not halt) on input $x\in\FS$ and auxiliary input $y\in\FIS$. $T$ is prefix-free if for all $x,s\in\FS$ with $s\,{\neq}\,\emptyset$, and $y\in\FIS$, either $T_y(x)\,{=}\perp$ or $T_y(xs)\,{=}\perp$ . The complexity of $x\in\FS$ with respect to $T_y$ is $\K_T(x|y)= \min\{\|p\|\,:\,T_y(p)=x\}$. 

There exists optimal for $\K$ prefix-free algorithm $U$, meaning that for all prefix-free algorithms $T$,  there exists $c_T\,{\in}\,\N$, where $\K_U(x|y)\leq \K_T(x|y)+c_T$ for all $x\,{\in}\,\FS$ and $y\,{\in}\,\FIS$. For example, one can take a universal prefix-free algorithm $U$, where for each prefix-free algorithm $T$, there exists  $t\in\FS$, with $U_y(tx)=T_y(x)$ for all $x\in\FS$ and $y\in\FIS$. The function $\K(x|y)$, defined to be $\K_U(x|y)$, is the Kolmogorov complexity of $x\in\FS$ relative to $y\in\FIS$. When we say that a universal Turing machine is relativized to an object, this means that an encoding of the object is provided to the universal Turing machine on an auxiliary tape. The universal probability of a number $a\in\N$ is $\m(a|y){= }\sum_z[\,U_y(z)=a]2^{-\|z\|}$. The coding theorem states $-\log \m(a|y)\eqa\K(a|y)$.

The chain rule for Kolmogorov complexity is $\K(x,y) \eqa \K(x)+\K(y|\langle x,\K(x)\rangle)$.  The mutual information in finite strings $x$ and $y$ relative to $z\in\FS$ is $\I(x\,{:}\,y\,{|}\,z)= \K(x|z)+\K(y|z)-\K(\langle x,y\rangle|z)\eqa\K(x|z)-\K(x|\langle y,\K(y|z),z\rangle)$.  

For computable probability measure $\lambda$ over $\IS$, $\K(\lambda)$ is the size of the shortest program that can compute $\lambda(x\IS)$, for all $x\in\FS$. For computable function $\delta:\IS\rightarrow\IS$, $\K(\delta)$ is the size of the shortest program on the input tape that outputs $\delta(\alpha)$ when $\alpha$ is on the auxiliary tape. The halting sequence $\mathcal{H}\in\IS$ is the infinite string where $\mathcal{H}[i]=[U(i)\neq \perp]$ for all $i\in\N$. The amount of information that $a\in\N$ has with $\mathcal{H}$ is denoted by $\I(a;\mathcal{H})=\K(a)-\K(a|\mathcal{H})$. 

The deficiency of randomness of $x$ with respect to elementary measure $Q$ and $v\in\N$ is $\d(x|Q,v)=\floor{-\log Q(x)}-\K(x|\langle Q\rangle,v)$. The stochasticity of $a\in\N$, conditional to $b\in\N$, is measured by $\Ks(a|b)=\min\{\K(Q|b)+3\log\max\{\d(a|Q,b),1\}\,{:}\,\textrm{$Q$ is an elementary probability measure}\}.$ We have $\Ks(a)=\Ks(a|\emptyset)$, with $\Ks(a|b)<\Ks(a)+O(\log\K(b))$.
\section{Sets of Numbers}
\label{sec:finiteset}
\begin{thr}
\label{thr:stochsub}
For computable probability $p$ over $\N$ and for $D\subset\FS$, $|D|=2^s$,  $m\in [0,s-1]$, there are $2^m$ elements $a\in D$, with $s-m < \d(a|p)+\Ks(D)+O(\log s+\log \K(p))$. 
\end{thr}
\begin{prf}
We relativize the universal Turing machine $U$ to $p$ and $s$ for the duration of the proof, which can be done as the theorem has precision $O(\log s)$. Let $Q$ be an elementary probability distribution that realizes $\Ks(D)$. Let $d=\max\{\d(D|Q),1\}$ be the deficiency of randomness of $D$ with respect to $Q$. We create an algorithm, that given $Q$, $s$ and $p$ produces $2^{s-1}$ sets $F_i\subseteq V$. We start with the first round. Suppose each element $a\in\N$ is selected independently with probability $cd2^{-s}$, where $c$ is a constant to be chosen later. The selected set is $F_1$, and $\E[p(F_1)]\leq cd2^{-s}$. Furthermore 
\begin{align*}
\E[Q(\{G:|G|=2^s,G\cap F_1=\emptyset\})]\leq \sum_GQ(G)(1-cd2^{-s})^{2^s}< e^{-cd}.
\end{align*}
 Thus a finite set $F_1$ can be chosen such that $p(F_1)\leq 2cd2^{-s}$ and $Q(\{G:|G|=2^s,G\cap F_1=\emptyset\}) \leq e^{1-cd}$. Now it must be that $D\cap F_1\neq \emptyset$. Otherwise, using the $Q$-test, $t(G)=[|G|=2^s, G\cap F_1=\emptyset]e^{cd-1}$, we have
\begin{align*}
\K(D|Q,d,c) & \lea -\log Q(D)-(\log e)cd\\
(\log e)cd & \lea -\log Q(D)-\K(D|Q) + \K(d,c) \\
(\log e)cd & \lea d + \K(d,c),
\end{align*}
which is a contradiction for large enough $c$ solely dependent on the universal Turing machine $U$. We make the relativization of $p$ and $s$ explicit for the following 3 equations. So there is an $a\in D\cap F_1$, where 
\begin{align*}
\K(a|p,s) &\lea -\log p(a) +\log d -s +\K(d|p,s)+\K(Q|p,s)\\
s&< \d(a|p)+\log d + \Ks(D|p,s)+ O(\log s)\\
s &< \d(a|p)+\Ks(D)+O(\log s).
\end{align*}
We define the construction of set $F_i$ given that the first $i-1$ rounds have already occurred. Let $F_{<i}=\bigcup_{j=1}^{i-1}F_j$. A set $G$ is eligible if $|G|=2^s$, and $|G\setminus F_{<i}| \geq 2^{s}-(i-1)$. Let $F_i$ be the random set where element $a\in\N$ is selected at random with probability $cd_i2^{-s}$, with $d_i = d\log i$. $\E[p(F_i)] \leq cd_i2^{-s}$. 
\begin{align*}
&\E[Q(\{ G : G\textrm{ is eligible and } (G\setminus F_{<i})\cap F_i=\emptyset\}) \\
\leq&  \sum_{\textrm{ eligible G}}Q(G)(1-c{d_i}2^{-s})^{2^s-(i-1)} \\
\leq& e^{-cd_i2^{-s}(2^s-(i-1))}\leq e^{-cd_i/2}.
\end{align*}
Thus a finite set $F_i$ can be chosen such that $p(F_i)\leq 2cd_i2^{-s}$ and $Q(\{G:G\textrm{ is eligible }, (G\setminus F_{<i})\cap F_i =\emptyset] \leq e^{-cd_i/2+1}$. It must be that on the rounds $i$ that $D$ is eligible, $(D\setminus F_{<i})\cap F_i\neq\emptyset$. Otherwise one can create a $Q$-test $t_i(G)=[G\textrm{ is eligible}, (G\setminus F_{<i})\cap F_i=\emptyset]e^{cd_i/2-1}$. Thus $t_i(D)\neq 0$ and
\begin{align*}
\K(D|Q,d_i,i,c) \lea -\log Q(D) - (\log e)cd_i/2\\
.5(\log e)cd\log i \lea -\log Q(D) - \K(D|Q) + \K(d_i,i,c)\\
.5(\log e)cd\log i \lea d+ \K(d,i,c).
\end{align*}
This is a contradiction for large enough $c$ dependent solely on the universal Turing machine $U$. We make the relativization of $p$ and $s$ explicit for the rest of the proof. Thus on rounds $i$ where $D$ is eligible, there exist an $a \in (D\setminus F_{<i})\cap F_i$, with
\begin{align}
\nonumber
\K(a|p,s) &\lea -\log p(a)+\log d_i -s +\K(d_i|p,s)+\K(i|p,s)+\K(Q|p,s)\\
\nonumber
s&< \d(a|p) +\log i+O(\log\log i)+\log d+\Ks(D|p.s)+O(\log s)\\
\label{eq:numlast}
s-\log i&< \d(a|p) +\Ks(D)+O(\log s+\log \K(p)).
\end{align}

On rounds $i$ in which $D$ is not eligible, then there exist a round $j<i$ where $|(D\setminus F_{<j})\cap F_j|>1$. And for all the elements in the intersection, a bound on their deficiency of randomness similar to Equation \ref{eq:numlast} can be proven.
\qed
\end{prf}

\begin{cor}
\label{cor:stochsub}
For computable probability $p$ over $\N$ and for $D\subset\FS$, $|D|=2^s$,  $s <\max_{a\in D}\d(a|p)+\Ks(D)+\K(s)+O(\log\K(p)+\log \K(s))$. 
\end{cor}
\section{Left-Total Machines}
\label{sec:LeftTotal}
\begin{figure}[h!]
	\begin{center}
		\includegraphics[width=0.4\columnwidth]{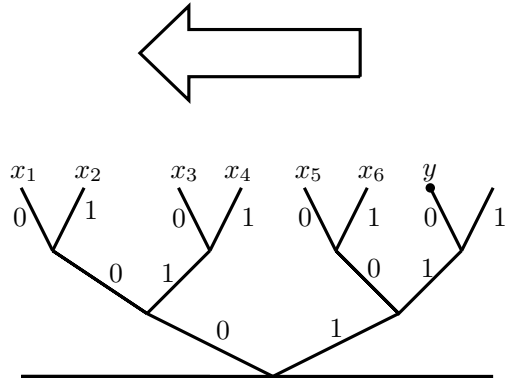}
		\caption{The above diagram represents the domain of a left total machine $T$ with the 0 bits branching to the left and the 1 bits branching to the right. For $i\in \{1,\dots,5\}$, $x_i\lhd x_{i+1}$ and $x_i\lhd y$. Assuming $T(y)$ halts, each $x_i$ is total. This also implies each $x_i^-$ is total as well.}
		\label{fig:LeftTotal}
	\end{center}
\end{figure}

We say $x\in\FS$ is total with respect to a machine if the machine halts on all sufficiently long extensions of $x$. More formally, $x$ is total with respect to $T_y$ for some $y\in\FIS$ if there exists a finite prefix free set of strings $Z\subset\FS$ where $\sum_{z\in Z}2^{-\|z\|}=1$ and $T_y(xz)\neq\perp$ for all $z\in Z$.  We say $\alpha\in\FIS$ is to the ``left'' of $\beta\in\FIS$, and use the notation $\alpha\lhd \beta$, if there exists $x\in\FS$ such that $x0\,{\sqsubseteq}\, \alpha$ and $x1\,{\sqsubseteq}\, \beta$. A machine $T$ is left-total if for all auxiliary strings $\alpha\in\FIS$ and for all $x,y\in\FS$ with $x\lhd y$, one has that $T_\alpha(y)\neq\perp$ implies that $x$ is total with respect to $T_\alpha$. An example  left-total machine can be seen in Figure \ref{fig:LeftTotal}.

For the remaining part of this paper, we can and will change the universal self delimiting machine $U$ into an optimal left-total machine $U'$. For a detailed explanation on how to construct a left-total universal Turing machine, we refer readers to \cite{Epstein21}. Without loss of generality, the complexity terms of this paper are defined with respect to the optimal left total machine $U$. For total string $b$, let $\B(b)=\max\{U(p):p\lhd b \textrm{ or } p\sqsubseteq b\}$ be the largest number produced by a program that extends $b$ or is to the left of $b$.

\section{Sets of Infinite Strings}

\begin{dff}[\cite{Levin74}]
\label{def:mutinf}
The mutual information between two infinite sequences $\alpha$ and $\beta$ is $$\I(\alpha:\beta|z)=\log\sum_{x,y\in\FS}\m(x|z,\alpha)\m(y|z,\alpha)2^{\I(x:y|z)}.$$
\end{dff}

\begin{lmm}[\cite{EpsteinLe11}]
	\label{lmm:StochH}
	For  $x\in\N$, $\Ks(x)\lel\I(x\,{;}\,\ch)$.
\end{lmm}

\begin{lmm}[\cite{Epstein21}]
	\label{lmm:totalString}
	If $b\in\FS$ is total and $b^-$ is not, and $x\in\FS$, \\
	then $\K(b)+\I(x;\mathcal{H}|b)\lel \I(x\,{;}\,\mathcal{H})+\K(b|\langle x,\|b\|\rangle)$.
\end{lmm}
\begin{lmm}[\cite{Epstein21}]
	\label{lmm:KTotalLength}
	If $b\in\FS$ is total and $b^-$ is not, and for $x\in \FS$, $\K(b|\langle x,|b\|\rangle)=O(1)$, then $\K(\|b\|)\lel 2\log\I(x;\ch)$.
\end{lmm}\newpage

\begin{thr}
\label{thr:subsetinf}
For computable probability $P$ over $\IS$ and  $Z\subset\IS$, $|Z|=2^s$, $m\in [0,s-1]$, there are $2^m$ elements $\alpha\in Z$, with $s-m< \D(\alpha|P)+\I(Z:\chi)+O(\log s +\log \I(Z:\chi)+\log \K(P))$. 
\end{thr}
\begin{prf}
The proof of this theorem follows closely in form to the proof of Theorem 5 in \cite{Epstein21}, except Theorem \ref{thr:stochsub} is referenced. Fix $m\in [0,s-1]$. Let $b$ be the shortest total string such that $|Z_{\B(b)}|=2^s$.Let $n=\B(b)$ and set $D=Z_{n}$. Let $p(x) = [\|x\|=n]P(\{\alpha : x\sqsubset \alpha\})$. Using Theorem \ref{thr:stochsub}, relativized to $b$, produces $2^m$ elements $F\subseteq D$ such that for $x\in F$, 
\begin{align*}
\K(x|b) &< -\log p(x)-s+m+\Ks(D|b)+O(\log s+\log \K(p|b)),\\
\K(x|b) &< -\log p(x)-s+m+\Ks(D|b)+O(\log s+\log \K(P)).
\end{align*}
Using Lemma \ref{lmm:StochH}, relativized to $b$,
\begin{align*}
\K(x|b) &< -\log p(x) -s +m+\I(D;\chi|b) + O(\log s+\log \I(D;\chi|b)+\log\K(P))\\
s{-}m &{<} \log (\m(x)/p(x)) {+}\K(b){+}\I(D;\chi|b){+} O(\log s {+}\log (\I(D;\chi|b){+}\K(b)){+}\log\K(P)).
\end{align*} 
By Lemma \ref{lmm:totalString},
\begin{align*}
s-m{<}& \log (\m(x)/p(x)) {+}\I(D;\chi){+}\K(b|D,\|b\|){+} \\
&{+}O(\log s{+}\log (\I(D;\chi){+}\K(b|D,\|b\|)){+}\log\K(P)).
\end{align*}
Since $D\subseteq \BT^{n}$, $\K(b|D,\|b\|)=O(1)$, as a program can output the leftmost total string $y$ of length $\|b\|$ such that $\B(y)$ is the length of the strings in $D$. So
\begin{align*}
s-m&< \log (\m(x)/p(x)) +\I(D;\ch)+O(\log s+\log \I(D;\chi)+\log \K(P)).
\end{align*}
We have that $\K(D|\langle Z\rangle)\lea \K(\|b\|)+\K(s)$, as $D$ is computable from $\langle Z\rangle$, $\|b\|$, and $s$. This is because $b$ is computable from its length, $s$, and $\langle Z\rangle$, and thus so is $D= Z_{n}$. By the Definition \ref{def:mutinf} of the mutual information between infinite sequences,
\begin{align}
\nonumber
\I(D;\chi) &\lea \I(\langle Z\rangle:\chi) + \K(D|\langle Z\rangle)\\
\nonumber
&\lea \I(\langle Z\rangle:\chi) + \K(\|b\|)+\K(s)\\
\label{eq:removeD}
&\lea \I(\langle Z\rangle:\chi) + 2\log\I(D;\chi)+\K(s)\\
\nonumber
&\lel \I(\langle Z\rangle:\chi) +\K(s),
\end{align}
where Equation \ref{eq:removeD} is due to Lemma \ref{lmm:KTotalLength}, noting $\K(b|D,\|b\|)=O(1)$. So there is an $\alpha \in Z$, $x\sqsubset \alpha$, with
\begin{align*}
s-m &< \log (\m(x)/p(x)) + \I(D;\chi)+O(\log s+\log \I(D;\chi)+\log \K(P))\\
s-m &< \D(\alpha|P) +\I(\langle Z\rangle:\chi)+O(\log s+\I(\langle Z\rangle:\chi)+\log \K(P)).
\end{align*}\qed
\end{prf}\newpage
\section{Outliers in Dynamics}
\label{sec:outdyn}
This section contains the main result of the paper, that dynamical systems will exhibit outliers. To prove this fact, Theorem \ref{thr:coninfo} will be leveraged, which details the conservation properties of the halting sequence $\chi$.

A dynamical system is defined by $(\lambda,\delta)$ where $\lambda$ is a probability over $\IS$ and $\delta:\IS\rightarrow\IS$ is continuous. For a continuous function $\delta:\IS\rightarrow\IS$, $\langle\delta\rangle$ is any infinite sequence $\delta'\in\IS$, such that if $\alpha\in\IS$ is on auxiliary tape of $U$ and $\delta'$ is on the input tape, $U$ outputs $\delta(\alpha)$ on the output tape, without halting. Similarly for arbitrary (i.e. uncomputable) probability measure $\lambda$ over $\IS$, $\langle \lambda\rangle$ is any infinite sequence $\lambda'$ such that if $x\in\FS$ is on the auxillary tape and $\lambda'$ is on the input tape of $U$, then $U$ outputs $\lambda(x\IS)$. For probability $\lambda$, $\I(\lambda:\chi)=\inf_{\langle\lambda\rangle}\I(\langle \lambda\rangle:\chi)$. We say that for dynamical system $(\lambda,\delta)$, $\I((\lambda,\delta):\chi)=\inf_{\langle \lambda\rangle,\langle \delta\rangle}\I(\langle \langle \lambda\rangle,\langle\delta\rangle\rangle:\chi)$.

\begin{thr}[\cite{Vereshchagin21,Levin74,Geiger12}]$ $\\
\vspace*{-0.5cm}
	\label{thr:coninfo}
	\begin{itemize}
 \item $\E_{\alpha\sim\lambda}\left[2^{\I(\alpha:\chi)}\right]\lem 2^{\I(\lambda:\chi)}$.
  \item $\I(f(\alpha):\chi)\lea \I(\alpha:\chi)+\K(f)$.
  \end{itemize}
\end{thr}

\begin{thr}[Outliers in Dynamics]
\label{thr:dynmut}
There exists $d\in \N$, where for computable probability $\mu$ over $\IS$ and dynamics $(\lambda,\delta)$, with $\I((\lambda,\delta):\chi)\neq\infty$, for $\lambda$-a.e. starting states $\alpha\in\IS$, there exists $s_\alpha\in\N$, where among the first $2^m$ states visited, for any $n<m$, there are at least $2^n$ states $\beta$ with $\D(\beta|\mu) > m-n-d\log m-s_\alpha$. Furthermore, for the smallest such $s_\alpha$, $\E_{\alpha\sim\lambda}\left[s_\alpha-O(\log s_\alpha)\right]<\I((\lambda,\delta):\chi)+O(\log \K(\mu))$.\end{thr}

\begin{prf}
 Fix a starting state $\alpha\in\IS$ and fix $d\in \N$. Assume $\alpha$ has property $A$, in which for all $s_\alpha\in\N$, there exists $m,n\in\N$, $m<n$, where the first $2^n$ states $Z\subset\IS$ visited has less than $2^m$ states $\beta\in Z$, with 
$$\D(\beta|\mu) > n - m - d\log n-s_\alpha.$$ 
Therefore, due to Theorem \ref{thr:subsetinf} there exists a state $\beta \in Z$, with $$\D(\beta|\mu) \leq n - m -d \log n-s_\alpha$$ and $$n-m<\D(\beta|\mu)+\I(Z:\chi)+O(\log\K(\mu)+\log n +\log\I(Z:\mathcal{H})).$$ 
Due to Theorem \ref{thr:coninfo}, we have 
$$\I(Z:\chi)\lea \I(\langle \alpha,\delta\rangle:\chi)+\K(n),$$ so
$$n-m<\D(\beta|\mu)+\I(\langle \alpha,\delta\rangle:\chi)+O(\log \K(\mu)+\log n +\log(\I(\langle\alpha,\delta\rangle:\mathcal{H})).$$ So $$n-m < n-m-d\log n-s_\alpha + \I(\langle \alpha,\delta\rangle:\chi) + O(\log\K(\mu)+\log n+\log\I(\langle\alpha,\delta\rangle:\chi)),$$ implying 
\begin{align*}
d\log n +s_\alpha&< \I(\langle\alpha,\delta\rangle:\chi)+ O(\log\K(\mu)+\log n+\log\I(\langle \alpha,\delta\rangle:\chi)).
\end{align*}
 Thus for large enough $d$, dependent solely on the universal Turing machine $U$, $\I(\langle\alpha,\delta\rangle : \chi)=\infty$. Let $\gamma(\langle\xi,\zeta\rangle) = \lambda(\xi)[\zeta=\langle\delta\rangle]$. By Theorem \ref{thr:coninfo}, $\E_{\alpha\sim\lambda}[2^{\I(\langle\alpha,\delta\rangle:\chi)}]=\E_{\xi\sim\gamma}[2^{\I(\xi:\chi)}]\lem 2^{\I(\gamma:\chi)}\lem 2^{\I(( \lambda,\delta):\chi)}<\infty$. Thus by Theorem \ref{thr:coninfo}, $\lambda$-a.e. states $\alpha$ do not have the property $A$.

By the reasoning above, the smallest such $s_\alpha$ has $s_\alpha \lel \I(\langle \alpha,\delta\rangle:\chi)+O(\log \K(\mu))$. So $s_\alpha - O(\log s_\alpha) < \I(\langle \alpha,\delta\rangle:\chi)+O(\log \K(\mu))$. So by Theorem \ref{thr:coninfo} and noting that $\alpha\mapsto c_\alpha$ is measurable, $\E_{\alpha\sim\lambda}\left[2^{s_\alpha}s_\alpha^{-O(1)}\right]\lem\E_{\alpha\sim\lambda}\left[2^{\I(\langle\alpha,\delta\rangle:\chi)+O(\log \K(\mu))}\right]\lem\E_{\xi\sim\gamma}\left[2^{\I(\xi:\chi)+O(\log\K(\mu))}\right]$ implies $\E_{\alpha\sim\lambda}\left[s_\alpha-O(\log s_\alpha)\right]< \I(\gamma:\chi)+O(\log \K(\mu))< \I(( \lambda,\delta):\chi)+O(\log \K(\mu))$.
\qed
\end{prf}
\begin{cor}[Computable Dynamics]
\label{cor:compdyn}
There exists $d\in \N$, where for computable probability $\mu$ over $\IS$ and computable dynamics $(\lambda,\delta)$, for $\lambda$-a.e. starting states $\alpha\in\IS$, there exists $s_\alpha\in\N$, where among the first $2^m$ states visited, for any $n<m$, there are at least $2^n$ states $\beta$ with $\D(\beta|\mu) > m-n-d\log m-s_\alpha$. Furthermore, for the smallest such $s_\alpha$, $\E_{\alpha\sim\lambda}\left[s_\alpha-O(\log s_\alpha)\right]< \K(\lambda,\delta)+O(\log \K(\mu))$.\end{cor}
This follows from Theorem \ref{thr:dynmut} and Theorem \ref{thr:coninfo}, where $\I((\lambda,\delta):\chi)\lea \I(0^\infty:\chi)+\K(\lambda,\delta)\lea\K(\lambda,\delta)$.
\section{Independence Postulate}
\label{sec:ip}

In this section, we show how the Independence Postulate can be used to explain outliers found in the physical world. The Independence Postulate (\textbf{IP}), \cite{Levin84,Levin13}, is an unprovable inequality on the information measure of two sequences. \textbf{IP} is a finitary Church Turing Thesis, postulating that certain infinite and \textit{finite} sequences cannot be found in nature, i.e. have high ``addresses''. One such example is finite prefixes of the halting problem. If a forbidden sequence can be found with a low address, then an ``information leak'' occurs.  L. A. Levin states \cite{Levin13}
\begin{quote}
\textit{The toolkit of our models may change (e.g., quantum amplitudes work somewhat differently than probabilities) but it is hard to expect new realistic primitives allowing such ``information leaks''.}
\end{quote}

The statement of the $\textbf{IP}$ is as follows.\\

\noindent\textbf{IP}\textit{: Let $\alpha$ be a sequence defined with an $n$-bit mathematical statement (e.g., in PA or set theory), and a sequence $\beta$ can be located in the physical world with a $k$-bit instruction set (e.g., ip-address). Then $\I(\alpha:\beta)<k+n+c$ for some small absolute constant $c$.}\\

We take $\I$ be the information term in Definition \ref{def:mutinf}. One consequence of \textbf{IP} is a finite version of the (physical) Church-Turing Thesis. \textbf{IP} says that the only finite sequences that can be found in nature (i.e. have short physical addresses) will have non-recursive descriptions that are equal in length to their recursive descriptions. This can be seen when \textbf{IP} is applied to the case when $\alpha=\beta$ is a finite sequence which has a non-recursive description of length $n$ that is much shorter than its recursive description $\K(\alpha)$, with $n\ll \K(\alpha)$. Let $k$ be the shortest physical address of $\alpha$. Then by \textbf{IP},
\begin{align}
\nonumber
\K(\alpha) \lea \I(\alpha:\alpha)  &\lea k + n +c\\
\label{eq:ipforbseq}
\K(\alpha)-n-c &\lea k.
\end{align}
Thus $k$ is large and $\alpha$ cannot be easily located in the physical world. The only sequences $\alpha$ with short physical addresses must have $n\approx\K(\alpha)$.

\subsection{Halting Sequence}
\label{sec:halting}
An example of the inequality in Equation \ref{eq:ipforbseq} is prefixes of the halting sequence. Let $H_n$ be the finite sequence that is the prefix of size $2^n$ of $\mathcal{H}$. It is well known that $\K(H_n)\in (n-O(1),n+\K(n)+O(1))$. The entire halting sequence $\mathcal{H}$ can be described in a mathematical statement of size equal to some small constant $c_{\mathrm{HM}}$. Each $n$ can be described using a program of size $\K(n)$. Therefore each $H_n$ can be defined by a mathematical statement of size $c_{\mathrm{HM}}+\K(n)$. So by \textbf{IP} applied to $\alpha=\beta=H_n$ where $k_n$ is the smallest physical address of $H_n$,
\begin{align*}
n \lea \K(H_n) \lea \I(H_n:H_n)&\lea c_{\mathrm{HM}}+\K(n)+c+k_n\\
n-\K(n)-(c_{\mathrm{HM}}+c) &\lea k_n.
\end{align*}

Therefore the prefixes of $\mathcal{H}$ do not exist in reality because, up to a small additive constant $(c_{\mathrm{HM}}+c+O(1))$, the  information content of $H_n$ is in the range $(n-O(1),n+\K(n)+O(1))$ but by \textbf{IP} the smallest physical address to reach $H_n$ is more than $n-\K(n)$.

\subsection{Peano Arithmetic}
\textbf{IP} can also be used in instances where $\alpha\neq\beta$, and one canonical example is to logic, and in particular Peano Arithmetic (PA). PA is a logic system that encodes statements of arithmetic through a set of initial axioms and a deduction system. G\"{o}del proved that PA is incomplete, in that there are well formed formulas in the language of PA which are true but are unprovable in PA. Suppose we order every well formed formula of PA and let the infinite sequence $L$ be defined such that its $i$th bit is 1 iff the $i$th formula of PA is true. Then $L$ is undecidable, in that there is no algorithm that can compute it. However G\"{o}del himself thought that there can be other means to produce true axioms of mathematics \cite{Godel61}:
\begin{quote}
\textit{
Namely, it turns out that in the systematic establishment of the axioms of mathematics, new axioms, which do not follow by formal logic from those previously established, again and again become evident. It is not at all excluded by the negative results mentioned earlier that nevertheless every clearly posed mathematical yes-or-no question is solvable in this way. For it is just this becoming evident of more and more new axioms on the basis of the meaning of the primitive notions that a machine cannot imitate.
}
\end{quote}
However, as detailed in \cite{Levin13}, \textbf{IP} forbids such information leaks. The sequence $L$ can be defined by a small mathematical formula of size $n$. Let $\beta$ be any source of information with a reasonably small physical address of size $k$, such as the contents of an entire mathematical library. Then by $\textbf{IP}$, with $\alpha=L$, this information source will have negligible shared information with $L$ (which encodes PA):
\begin{align*}
\I(\beta:L)&< k+n+ c.
\end{align*}

\subsection{Inductive Inference}

The deficiency of randomness $\d(a|p)$ is a score that refutes the hypothesis that $a \in \N$ was generated by the model $p$, which is a probability measure over $\N$. If $p=\m$, defined in Section \ref{sec:conv}, then for all $a \in \N$, $\d(a|\m)=O(1)$.  Thus there is no refutation to the statement: “$a$ was generated by $\m$”. This implies $\m$ acts as a universal ``a-priori'' distribution over $\N$. 

In inductive inference, the goal is to describe the environment from a set of candidates, $D$.  In its essence, this is the process of selecting an element $x \in \FS$ from a set $D \subseteq \FS$. The set $D$ can be seen as the group of (very long) descriptions of the environment consistent with an observation. Let $\m(D) = \sum_{x \in D} \m(x)$, be the total a-priori probability of the members of $D$. In \cite{EpsteinLe11}, it was shown that the a-priori probability of a non-exotic set of candidates $D$ is focused on a single observation $x \in D$, with
$$
\min_{x \in D} -\log \m(x) \lel -\log \m(D) + \I(D;H).
$$

We can apply \textbf{IP} to this inequality, and for this section only, we use $\I(x;\chi) = \K(x) - \K(x|\chi)$ as the information term in \textbf{IP}. Let $k$ be the size of a physical address for the candidates $D$. Let $c_\mathrm{HM}$ be the smallest mathematical description of the halting problem. By \textbf{IP},
$$
(\min_{x \in D}-\log \m(x)) + \log \m(D) \lessapprox k +  c_\mathrm{HM} + c.
$$

Thus observations in the physical world, (i.e. sets with small addresses) will have explanations (i.e. elements of the set) with a-priori probability concentrated on a single explanation. Due to the coding theorem, this element is equal to $\arg \min_{x \in D}\K(x)$, the simplest explanation of the given observation.  

\subsection{Outliers}
 This section shows that observable sequences in the physical world will have emergent outliers. The results of Section \ref{sec:finiteset}, along with Lemma \ref{lmm:StochH} show us that finite sets of numbers will have outliers or large mutual information with the halting sequence. This can easily applied to individual infinite sequences of numbers, as shown in Theorem \ref{thr:infpref}. For $\tau\in\N^\N$, $\langle\tau\rangle=\langle \tau[1]\rangle\langle \tau[2]\rangle\langle \tau[3]\rangle\dots$ Let $\tau(n)$ be the first $2^n$ unique numbers found in $\tau$. The sequence $\tau$ is assumed to have an infinite amount of unique numbers, and represents a series of observations.
 \begin{thr}
\label{thr:infpref}
For probability $p$ over $\N$ and $\tau\in\N^\N$, let $s_{\tau,p}=\sup_n\left( n - 3\K(n)-\max_{a\in\tau(n)}\d(a|p)\right)$. Then $s_{\tau,p}\lel \I(\langle \tau\rangle:\ch)+O(\log\K(p))$.
\end{thr}
\begin{prf} By Corollary \ref{cor:stochsub}, Lemma \ref{lmm:StochH},  and the fact that $\I(x;\chi)\lea \I(\alpha:\chi)+\K(x|\alpha)$,
\begin{align*}
n&< \max_{a\in \tau(n)} \d(a|p) +\I(\tau(n);\chi)+\K(n)+O(\log(\I(\tau(n);\chi)+\K(p)+\K(n)))\\
n&< \max_{a\in \tau(n)} \d(a|p) +2\K(n)+\I(\langle\tau\rangle:\chi)+O(\log(\I(\langle\tau\rangle:\chi)+\K(p)+\K(n)))\\
n&- 3\K(n) - \max_{a\in \tau(n)} \d(a|p)\lel\I(\langle\tau\rangle:\chi)+O(\log \K(p)).
\end{align*}\qed
\end{prf}\\
 Let $k$ be the physical address of an infinite sequence of numbers $\tau$. As defined above, the halting sequence $\chi$ can be described in a mathematical statement of size equal to some small constant $c_{\mathrm{HM}}$. Then by Theorem \ref{thr:infpref}, for some sequence $\tau\in\N^\N$ and probability $p$, \textbf{IP} states
\begin{align*}
s_{\tau,p}\lel \I(\langle\tau\rangle:\chi)+\K(p) \lel k + c_{\mathrm{HM}}+c+O(\log \K(p)).
\end{align*}
Thus sequences $\tau$ with large $s_\tau$, as defined in Theorem \ref{thr:infpref}, will have large physical address. Thus it is hard to find physical sequences which do not have large outliers, and completely impossible to find sequences with no outliers. As the complexity of the probability $p$ in the randomness deficiency term increases, the bounds loosen. Theorem \ref{thr:infpref} can be strengthened to characterize subsets of initial subsequences, in similar form to Theorem \ref{thr:dynmut}. One can also use \textbf{IP} to characterize sequences of infinite sequences or reals.
\section{Uncomputable Sampling Methods}
As mentioned in the introduction, a sampling method $A$ takes in a number $n$, and outputs, with probability 1, $n$ unique numbers or infinite sequences. In \cite{Epstein21}, it was shown that computable sampling methods produce outliers. In \cite{Epstein22}, the bounds were improved and results were proven about sampling methods that can possibly not halt with positive probability. In this section, we show that discrete and continuous sampling methods that are uncomputable but whose encodings has finite information with the halting sequence will produce outliers.

An uncomputable sampling method is a function $A:D\rightarrow \N^*$, where $D\subset \N\times\FS$, and $D_n$ is a prefix free set for all $n\in\N$. Furthermore $\sum_{x\in D_n}2^{-\|x\|}=1$ and for all $x\in D_n$, $|A(n,x)|=n$. An encoding $\langle A\rangle$ of a discrete uncomputable sampling method $A$, is any infinite sequence, such that if it is on the input tape of the universal Turing machine $U$, and $\langle n,\omega\rangle$ is on the auxiliary tape, $U$ outputs $n$ elements using random seed $x\sqsubset\omega$, $x\in D_n$, and then halts. $\I(A:\chi) = \inf_{\langle A\rangle}\I(\langle A\rangle:\chi)$.

\begin{thr}
\label{thr:uncompsamp}
For discrete (possibly uncomputable) sampling method $A$, computable probability $p$ over $\N$, $\Pr_{D\sim A(2^n)}[n-k>\max_{a \in D}\d(a|p)] < 2^{-k+\I(A:\chi) +O(\log n)+c_p}$.
\end{thr}
\begin{prf}
Fix any $D\subset\N$, $|D|=2^n$, and let $d_D=\max_{a \in D}\d(a|p)$. By Corollary \ref{cor:stochsub} and Lemma \ref{lmm:StochH}, we have
$$
n < d_D+ \I(D;\ch) + O(\log(n)+\log(\I(D;\ch))) + c_p.
$$
So $ n - d_D- O(\log n) < \I(D;\ch)+c_p$. So
$$
\E_{D\sim A(2^n)}\left[2^{n-d_D-O(\log n)}\right]<\E_{D\sim A(2^n)}\left[2^{\I(D;\ch)}\right]2^{c_p}.
$$
Let $\gamma_n$ be a probability measure over $\IS$, where $\gamma_n(\langle D\rangle0^\infty)=\Pr(A(2^n)=D)$. From Theorem \ref{thr:coninfo},
\begin{align*}
&\E_{D\sim A(2^n)}\left[2^{n-d_D}\right]\\
<&\E_{D\sim A(2^n)}\left[2^{\I(\langle D\rangle0^\infty:\chi)}\right]2^{O(\log n)+c_p}\\
<&\E_{\alpha\sim \gamma_n}\left[2^{\I(\alpha:\chi)}\right]2^{O(\log n)+c_p}\\
<&2^{\I(\langle\gamma_n\rangle:\ch)+O(\log n)+c_p}\\
<&2^{\I(\langle n,A\rangle:\ch)+O(\log n)+c_p}\\
<&2^{\I(A:\chi)+O(\log n)+c_p}.
\end{align*}
Thus we get the theorem statement, with
$$
\Pr_{D\sim A(2^n)}\left[n-k>\max_{a\in D}\d(a|p)\right]<2^{-k+\I(A:\ch)+O(\log n)+c_p}.
$$
\qed
\end{prf}

Better bounds than $O(\log n)$ can be achieved at the cost of complicating the proof.  

An uncomputable continuous sampling method is a continuous function $D\rightarrow {\left(\IS\right)}^*$ where $D\subseteq \N\times\IS$ and $\mathcal{U}(D_n)=1$, where $\mathcal{U}$ is the uniform probability. If $\alpha\in D_n$, then $A(n,\alpha)\in {\left(\IS\right)}^n$. The encoding of a continuous sampling method $A$, is an infinite sequence $\langle A\rangle$ such that if $\langle n,\omega\rangle$ is on the auxilliary tape, the universal Turing machine $U$ outputes $A(n,\omega)$ on the output tape. The output is encoded as:

$$\beta_1[1]\beta_2[1]\dots\beta_n[1]\beta_1[2]\dots$$

Furthermore, for continuous sampling method $A$, $\I(A:\ch)=\inf_{\langle A\rangle}\I(\langle A\rangle:\ch)$.
\begin{thr}
For (possibly uncomputable) continuous sampling method $A$, computable probability $P$ over $\IS$, $\Pr_{D\sim A(2^n)}[n-k>\max_{\alpha \in D}\D(\alpha|P)] < 2^{-k+\I(A:\ch)+O(\log n)+c_P}$.
\end{thr}

The proof follows almost identically to that of Theorem \ref{thr:uncompsamp}. We leave the details to the reader.
\label{sec:unsamp}
\section{Discussion}
The paper proves results for discrete time dynamical systems. An open problem is how outliers manifest in continuous time dynamical systems. Another avenue of research involves the connections of outliers in dynamical systems to thermodynamics. A good starting point for research into this area would be the paper \cite{Gacs94}, and the general research direction would be to prove properties of algorithmic fine grained entropy instead of randomness deficiency. The results in the paper is over the Cantor space. This sets the stage to characterize properties of computable dynamics over general spaces. Another application to \textbf{IP} is complete extensions of partial predicates and their relation to statistical learning theory, as discussed in \cite{Epstein21}.

\end{document}